%% file: main.tex
\documentclass[10pt,twocolumn,letterpaper]{article}

\makeatletter
\@namedef{ver@everyshi.sty}{}
\makeatother

\usepackage[pagenumbers]{cvpr} % To force page numbers, e.g. for an arXiv version

\usepackage{graphicx}
\usepackage{amsmath}
\usepackage{amssymb}
\usepackage{xcolor}
\usepackage{mathtools}

\input{macros}

\ifarxiv
\usepackage{hyperref}
\fi

\usepackage[capitalize]{cleveref}
\crefname{section}{Sec.}{Secs.}
\Crefname{section}{Section}{Sections}
\Crefname{table}{Table}{Tables}
\crefname{table}{Tab.}{Tabs.}

\def\cvprPaperID{23} % *** Enter the CVPR Paper ID here
\def\confName{CVPR}
\def\confYear{2024}

\begin{document}

% \title{LoopDraw: an Autoregressive Loop Primitive for Shape Synthesis and Editing}
% \title{LoopDraw: an Autoregressive Model for Loop-Based \\ Shape Synthesis and Editing}
\title{LoopDraw: a Loop-Based Autoregressive Model for Shape Synthesis and Editing}

\author{Nam Anh Dinh\\
University of Chicago\\
% Chicago, IL\\
{\tt\small namanh@uchicago.edu}
\and
Haochen Wang\\
TTI-Chicago\\
{\tt\small whc@ttic.edu}
\and
Greg Shakhnarovich\\
TTI-Chicago\\
{\tt\small greg@ttic.edu}
\and
Rana Hanocka\\
University of Chicago\\
{\tt\small ranahanocka@uchicago.edu}
}

% teaser with env stuff from cvpr23 and content (only thing changed was the includegraphics source, a prerendered pdf now rather than tikz)

\twocolumn[{%
\renewcommand\twocolumn[1][]{#1}%
\maketitle
\begin{center}
    \centering
    \newcommand{\transplant}{\color[rgb]{1,0.2,0.1}}
    \newcommand{\oldloop}{\color[rgb]{0.6,0.6,0.6}}
    \newcommand{\newloop}{\color[rgb]{1,0.75,0.79}}
    \captionsetup{type=figure}
    \includegraphics{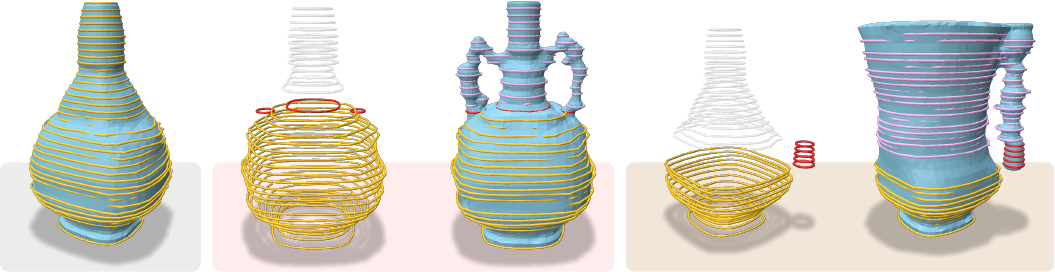}
    \caption{\ourmethod{} represents shapes as a sequence of loops. As the loop sequence unfolds, we {\transplant \textbf{modify a few loops}} on-the-fly during autoregressive decoding, resulting in structural shape modifications. In place of the {\oldloop \textbf{(original) subsequent loops}} of the shape, the autoregressive generation produces {\newloop \textbf{new loops}}.
    }
    \label{fig:teaser}
\end{center}%
}]

%%%%%%%%% ABSTRACT
\begin{abstract}
There is no settled universal 3D representation for geometry with many alternatives such as point clouds, meshes, implicit functions, and voxels to name a few. In this work, we present a new, compelling alternative for representing shapes using a sequence of cross-sectional closed loops. The loops across all planes form an organizational hierarchy which we leverage for autoregressive shape synthesis and editing. Loops are a non-local description of the underlying shape, as simple loop manipulations (such as shifts) result in significant structural changes to the geometry. This is in contrast to manipulating local primitives such as points in a point cloud or a triangle in a triangle mesh. We further demonstrate that loops are intuitive and natural primitive for analyzing and editing shapes, both computationally and for users.
\ifarxiv
See our project page at \url{https://threedle.github.io/LoopDraw}.
\fi
\end{abstract}

%%%%%%%%% BODY TEXT

% \input{01_intro}
% \input{02_relatedworks}
% \input{04_method}
% \input{05_experiments}
% \input{06_conclusion}
% \input{08_ack}

\input{01_intro}

\input{02_relatedworks}

\input{04_method}

\input{05_experiments_old}

\input{06_conclusion}

% for arxiv
\ifarxiv
\input{08_ack}
\fi

% \clearpage
% \clearpage
%%%%%%%%% REFERENCES
{\small
\bibliographystyle{ieee_fullname}
\bibliography{bibs}
}

% \clearpage
% \input{07_appendix}

\end{document}

%% file: macros.tex
\usepackage{amsmath}
\usepackage{graphicx}
\usepackage{multirow}
\usepackage{booktabs}

\definecolor{amber}{rgb}{1.0, 0.75, 0.0}
\definecolor{applegreen}{rgb}{0.55, 0.71, 0.0}
\definecolor{darkgoldenrod}{rgb}{0.72, 0.53, 0.04}
\definecolor{firebrick}{rgb}{0.7, 0.13, 0.13}

\newcommand{\ourmethod}{LoopDraw}

\newif\ifdraft
\draftfalse

\ifdraft
\newcommand{\nmc}[1]{{\color{red}[\textbf{Nam Anh:} #1]}}
\newcommand{\rhc}[1]{{\color{blue}[\textbf{Rana:} #1]}}
\newcommand{\whc}[1]{{\color{magenta}[\textbf{Haochen:} #1]}}
\newcommand{\gs}[1]{{\color{green!50!black}[\textbf{GS:} #1]}}
\newcommand{\rh}[1]{{\color{blue}#1}}
\newcommand{\nm}[1]{{\color{red}#1}}
\else
\newcommand{\nmc}[1]{}
\newcommand{\rhc}[1]{}
\newcommand{\whc}[1]{}
\newcommand{\gs}[1]{}
\newcommand{\rh}[1]{{\color{black}#1}}
\newcommand{\nm}[1]{{\color{black}#1}}
\fi

\newif\ifarxiv
\arxivtrue
\def\hlinewd#1{%
\noalign{\ifnum0=`}\fi\hrule \@height #1 %
\futurelet\reserved@a\@xhline} 

\usepackage{xcolor}

%% file: 01_intro.tex
\section{Introduction}
\rhc{TODO: no other loop editing works}
Applying deep learning models on irregular data, particularly shapes and geometry (in contrast to regular data such as pixel grids in images), remains challenging.
This is in part due to an underlying open question in geometry processing: how should shapes be represented? There is no settled universal representation among the many contenders, which include point clouds~\cite{qi2017pointnet}, meshes~\cite{hanocka2019meshcnn}, voxels~\cite{wu20153d}, and implicit functions~\cite{park2019deepsdf}, to name a few.

We present a \rh{new, compelling neural} alternative for representing shapes using a \textit{loop primitive}.
A set of loops gives a sparse, yet informative portrayal of the underlying surface. We posit that loops have unique advantages compared to existing representations, allowing for intuitive user-editing, interpretable visual inspection, shape encoding, and better informed-reconstruction. 

\rh{Representing 3D surfaces using planar cross-sectional slices has a long history in computer graphics, medical imaging, and geographical systems~\cite{bermano2011online, barequet2009reconstruction, kehtarnavaz1988syntactic, boissonnat2007shape, zou2015topology}. Each planar cross-section contains an arbitrary number of loops, and each loop is a closed two-dimensional contour. This work introduces \emph{\ourmethod{}}, an autoregressive model for synthesizing 3D shapes based on cross-sectional loop sequences. The loops across all planes form an organizational hierarchy that we leverage for neural shape synthesis and editing.}
The loop sequence data both \textit{describes} the underlying surface, and \textit{prescribes} a shape in the way it is reconstructed into a mesh.  

We show that this loop representation, encoded as sequence data, presents an effective neural primitive for a generative model of 3D shapes via a transformer-based variational autoencoder (VAE). Meshes are first sliced to yield loop data, which is then used to train an encoder and a decoder with a latent space bottleneck (see \cref{fig:overview}). The latent space smoothness offered by the VAE architecture and the intuitiveness and editability of loop data together form a powerful system for the controllable generation and refinement of 3D shapes. Tasks enabled by this framework include latent space sampling and interpolation between shapes, as well as manual loop editing during autoregressive decoding. This mid-decoding intervention allows meaningful, interpretable manipulation of latent-decoded shapes by manual modification of loops (scaling, shifting, deformations); it also enables shape blending and completion by inserting loops transplanted from a different shape.  

Loops bound the outer contours of the shape, enabling minimal edits to cause larger-scale effects to the resultant geometry. 
We opt to represent geometry as a sequence of loops learned with an autoregressive decoder. Manipulating predicted loops during autoregressive decoding yields cascading effects on remaining loops to adapt to the local edit (see \cref{fig:teaser,fig:morph-to-mug,fig:handle-transplant,fig:square-injection,fig:sofa-rectangle-injection,fig:vase-twoedit}). This property illustrates how loops are a \textit{non-local} description of the underlying shape. By contrast, editing other primitive representations tends to produce overly local effects, on the scale of individual triangles (in meshes), voxel cells (in voxel grids), or points (in point clouds). We may also contrast our approach with implicit neural representations which, while effective at reconstruction tasks, are not conducive to manual control post-generation~\cite{zekun2020dualsdf}. 

Loops are an \textit{explicit} representation of the underlying 3D surface. This enables accurately portraying sharp corners and high-frequency features of a surface (\cref{fig:teaser,fig:square-injection,fig:random_samples}). Moreover, it is straightforward to obtain from closed planar loops the inside/outside information, necessary for surface reconstruction~\cite{dipole}, in post-process. In contrast, point clouds require a consistent normal orientation for reconstruction, a notoriously difficult problem in geometry processing~\cite{dipole}. Lastly, loops are a native representation for 3D processing applications utilizing tomography (e.g. radio-imaging in biology and geology); our representation and system would offer a low-friction way to apply 3D generative modeling to these data domains~\cite{zou2015topology, fang2022topocut}.

In summary, we present a loop primitive as a novel neural representation of 3D shapes. Loops accurately portray the underlying shape contours, and contain non-local surface information. Our architecture is designed to autoregressively encode and decode a hierarchical set of loops, using a transformer-based variational autoencoder. We also demonstrate that loops are intuitive and natural primitive for analyzing and editing shapes, both computationally and directly by human intervention.

%% file: 02_relatedworks.tex
\begin{figure}[b]
    \centering
    % can't use overpic! here is the prerendered pdf with fig-renderer-main.tex
    \includegraphics{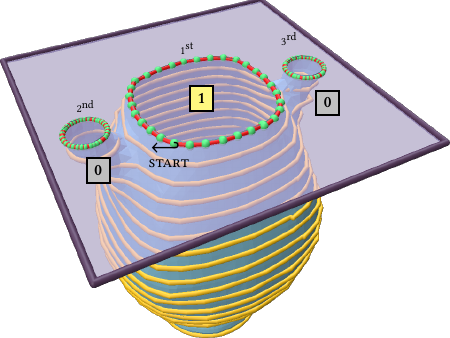}
     \caption{Closeup of a slice plane. In this example, the plane intersects the mesh at three closed loops, defining three time steps in our sequence data. Each loop has an associated level-up binary flag (shown as $\mathbf 0$ or $\mathbf 1$); the loop with a $\mathbf 1$ flag introduces this slice plane and comes first in the sequence data, followed by the other loops with a flag of $\mathbf 0$.}
    \label{fig:cross-section-closeup}
\end{figure}

\section{Related Work}

\begin{figure*}
    % 2024-01-20: originally at the top of method.tex, but moved here so that page 2 has the figure up
    \centering
    \includegraphics[width=\textwidth]{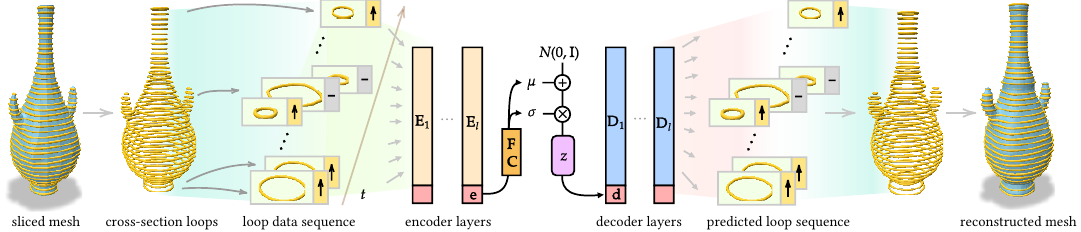}
    \caption{\ourmethod{} overview. Starting with an initial mesh, we extract a sequence of cross-sectional loops over a progression of planes. The entire sequence of loops (2D polygons with a level-up flag), augmented with a special token $\mathbf{e}$, is encoded through a series of $l$ transformer layers, producing features $[E_1,\mathbf{e}_1],\dots, [E_l,\mathbf{e}_l]$. The special token $\mathbf{e}_l$ in the last layer's output is used as an aggregate encoder embedding that is then mapped to the parameters of a latent distribution. The latent vector $\mathbf{z}$ is projected to the start-of-sequence embedding $\mathbf{d}_1$ for the decoder, in which $l$ transformer layers produce $D_1 \dots D_l$, which is trained to reconstruct the original input loop sequence. The reconstruction loss is the L2 distance (and cross-entropy for the binary flag) between the predicted and ground truth loop data, and a KL divergence term on the latent space parameters. The network-predicted loops are reconstructed to obtain a 3D mesh. In generation mode, we sample $\mathbf z$ from a standard Gaussian, and generate the sequence autoregressively.} 
    \label{fig:overview}
\end{figure*}

Our approach joins a number of other works in the field of shape generation, specifically those that leverage sequences. Among autoregressive sequence-based generative frameworks,  PolyGen~\cite{nash2020polygen} generates a sequence of mesh vertices and then learns to connect them together. MeshGPT~\cite{siddiqui2023meshgpt} models shapes as a sequence of triangles using language-model-style autoregressive decoding from a learned quantized latent vocabulary of faces. 
Other works with diverse approaches to the sequence-based 3D shape modeling problem include DeepCAD~\cite{wu2021deepcad}. We take inspiration from SketchRNN \cite{ha2018SketchRNN} and their neural representation of sketch strokes in the 2D vector graphics context. 
To our knowledge,
our work is the first technique to build a neural network for analyzing and synthesizing cross-sectional loops for shape generation and editing. \rh{This unique representation facilitates a method for achieving structural shape manipulations through simple and intuitive loop edits (such as geometric transformations) that imply broader changes to the resultant geometry.}

Distinct from the autoregressive approach, there have been methods modeling explicit surfaces using other representations such as patches \cite{groueix2018papier}, voxel grids \cite{sella2023vox, sanghi2023clipsculptor}), tetrahedral grids \cite{gao2022tetgan}, manifold surfaces via neural ODEs \cite{gupta2020neural}, programs composing together learned primitives \cite{jones2023ShapeCoder}, and most recently triangle soups as implemented in PolyDiff~\cite{alliegro2023polydiff}, a denoising diffusion model on triangle soups.
The explicit mesh representation has also been used for synthesizing fixed-topology modifications and distribution learning of shapes, such as geometric texture synthesis~\cite{Hertz2020geometric}, morphing and interpolation~\cite{maesumi2023explore}, text-based deformation~\cite{gao2023textdeformer}, and applications in analysis, reconstruction, and transmission~\cite{Liu:Subdivision:2020, uy2021joint3dretrievaldeform, point2mesh, chen2023neuralprogressivemeshes}. As an alternative to generating shapes with fixed-topology, some techniques produce individual parts (e.g. meshes or cuboids) that are stitched together~\cite{yang2022dsg, shapeassembly}. \rh{Several techniques proposed generative techniques for point clouds~\cite{ShapeGF,nakayama2023difffacto}.}
Editing these primitive representations (such as points, voxels, triangles, or volumetric cells) produce highly local effects. By contrast, loops are \textit{non-local}, where simple geometric edits resulting in broad changes to the final shape. Our use of loops as a neural primitive takes inspiration from classic works that utilize cross-sectional planar slices~\cite{zou2015topology, zeng2008topology, bermano2011online}. \rh{Moreover, three dimensional (non-planar) loops are also related to quad-meshing and estimating sharp feature lines in 3D~\cite{campen2012dual, zhuang2014anisotropic}.}

Departing from explicit representations, implicit field-based methods, as well as hybrids with explicit elements, are also well-represented in the neural shape modeling literature~\cite{chen2018imnet, park2019deepsdf, zekun2020dualsdf, mescheder2019occupancy, kleineberg2020shapegan, zhang20223dilg, erkoc2023hyperdiffusion, zhang20233dshape2vecset, Tertikas2023partnerf}. Neural fields as well as the new Gaussian splatting representation~\cite{kerbl3Dgaussians}, with their ability to fit high-fidelity visuals from images, have given rise to a large body of techniques for text-to-3D generation and 2D diffusion-based 3D optimizations.~\cite{sjc}~\cite{poole2022dreamfusion,wang2023prolificdreamer,tsalicoglou2024textmesh} While these methods enable a variety of editing modalities for implicit shape representations, we differ by presenting an explicit editing target via loops, being concrete explicit geometric objects that can be directly manipulated.

%% file: 04_method.tex
\section{Method}

In this section, we describe the details of our framework (see illustration in \cref{fig:overview}). We define shapes using a sequence of closed cross-sectional loops (\cref{sec:loop-repr}) and build a generative architecture for learning distributions of shapes using this representation (\cref{sec:enc-dec-arch}).

\subsection{Loop Representation}
\label{sec:loop-repr}
Our loop-based representation of a 3D shape is defined by a sequence of planar cross-sectional contours. We slice along one axis of the shape, using an evenly spaced list of planes $P$. For each plane $p_i \in P$, we extract a set of loops $S$. 
The cross-sections with each slice plane $p_i$ are stored in $S$ as polylines, and we require all meshes used in this work to be manifold (in the areas that intersect the chosen slice planes); in other words, we enforce all loops in $S$ to be closed loops. After slicing, we (uniformly) resample the loops to contain $N$ (in our case $32$) vertices. Each closed loop is one time step (i.e., token) in the sequence $S$ along the slicing axis.

A plane is defined by a point $\mathbf o_i \in \mathbb R^3$ and a normal vector $\vec{n}_i \in \mathbb R^3$.
Each plane $p_i$ in the list $P$ is encoded with these two vectors alongside
two orthonormal vectors $\vec {x}_i$ and $\vec{y}_i$ which lie on the surface of the plane.
Together these create a local coordinate system based on $p_i$ (with origin $\mathbf o_i$, $z$-axis $\vec n_i$, and $x-$ and $y-$ axes $\vec x_i$ and $\vec y_i$). Since our loops are planar (two-dimensional), each vertex of each loop on the plane $p_i$ can be represented with $\mathbb R^2$ coordinates with respect to this local system.

Even though each time step is one closed loop, each plane does not necessarily intersect the mesh at just one closed loop. 
We handle arbitrary number of per-plane loops by pairing each loop $s_t$ in $S$ (a closed polyline)  with a binary \textit{level-up} flag 
$u_t$.

This scheme implicitly records which plane each closed loop belongs to; the correspondence can be recovered by iterating across $S=([s_1,u_1],\ldots,[s_T,u_T])$ keeping track of the current plane $p_i$ in $P$ (initializing the state at $p_0$ which corresponds to \textit{no-plane}): for each time step $t$, the loop described in $s_t$ is assigned to the current $p_i$ unless a level-up flag $u_t=1$, in which case we update $p_i := p_{i+1}$ and assign the loop in $s_t$ to the new $p_i$. 
Thus, a slice plane having multiple cross-section polygons would accordingly be represented as multiple time steps in $S$, the first of which has a \textit{level-up} flag set to $1$ and the rest having it set to $0$. \gs{a tad repetitive} See \cref{fig:cross-section-closeup} for an illustration.

The per-loop polyline data is represented as a flattened vector of $N$ coordinates $s_t=\left[x_1, y_1,\dots,x_N, y_N\right]$ with respect to the coordinate system on the plane $p_i$ to which the loop belongs. We impose a canonical ordering for this array of $N$ points such that $(x_1, y_1)$ is the loop point with minimum $(x+y)$, and the rest of the $(x,y)$ loop points will be arranged in clockwise order using their 2D coordinates on the plane coordinate system. Not all loops are convex, with a consistent clockwise orientation throughout, so this enforcement is heuristically based on the orientation of the first three points of a loop.

\paragraph{Reconstructing a mesh from loops}
There exists methods to reconstruct meshes directly from contour loops \cite{zou2015topology}. Alternatively, because our loops are polylines and not unordered point sets, we may sample a point cloud and estimate surface normals directly from loops at arbitrary resolution (\cref{fig:normal-estimation}). This enables the option of reconstructing a mesh from an oriented point cloud sampled from loops.

\subsection{Network Architecture}
\label{sec:enc-dec-arch}
We model distributions of shapes represented as a sequence of loops using a variational autoencoder (VAE) \cite{kingma2013vae}. This choice is motivated by the observation that an autoencoder design would allow for a smooth latent space of shapes. This  complements the interpretability and ease of editing of the loop data itself, together enabling highly controllable, intuitive, and transparent generation and manipulation of loops and shapes.
Our architecture leverages transformers~\cite{vaswani2017transformer} as the backbone for the encoder and decoder networks.

\paragraph{Encoder} The input to the encoder is a full loop data sequence of variable length, and the output is a fixed-size aggregate embedding that summarizes the shape based on this sequence. Specifically, the input is the sequence $S$ of loop parameters (as described in \cref{sec:loop-repr}) plus a sinusoidal positional encoding.

To obtain this aggregate summary embedding, we insert a special token $\mathbf e$ (initialized to be $\mathbf 0$) at the beginning of the input sequence. This is encoded through the transformer layers alongside the rest of the sequence. 
Then, we extract the output embedding from the final layer only at that first token slot $\mathbf e$ (corresponding to the special token appended to the input) (\cref{fig:overview}).
This $\mathbf e$ is interpreted as an aggregate fixed-size embedding that describes the entire shape. The aggregate embedding is passed through MLPs to predict the parameters $(\mu, \log \sigma^2)$ of a distribution in $N_z$-dimensional latent space. Using the reparameterization trick, we use these parameters to sample a latent code $\mathbf z$ for the decoder.

\begin{figure}[b]
    \centering
    \includegraphics[width=\columnwidth]{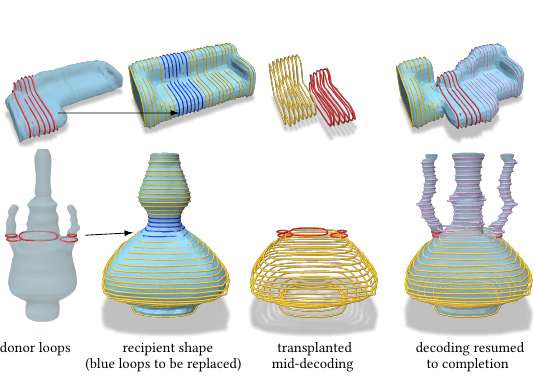}    
    ~
    
    \caption{Loop transplantation. By taking new loops (e.g., handles) hand-drawn or extracted from a donor shape, manually fixing them during autoregressive decoding of a ``recipient shape'' latent code, then decoding the rest with the edit present, we obtain adaptive shape continuations that best explain this manual insertion. In the upper example, the model smoothly completes a novel sofa given a manually-added protrusion from an L-shaped donor sample. In the lower example, the model completes the appropriate handles for the original handle-less vase. \rhc{would love another row of this :) }}
    \label{fig:handle-transplant}

\end{figure}

\paragraph{Decoder}
During training, the decoder network receives as input 1) a latent code $\mathbf z$ of size $N_z$, and 2) the training sequence shifted right one time step (i.e. with an empty start-of-sequence token.) (This data sequence is only provided at training time; at inference time, this is provided from previous output autoregressively.)
The latent code $\mathbf z$ is passed through an MLP with final $\tanh$ activation to create the start-of-sequence embedding $\mathbf d$ for the decoder, effectively making a learned start-of-sequence embedding conditioned on the latent code. Throughout decoding, this $\mathbf z$-conditioned start embedding is attended to, along with all subsequent time steps through self-attention, conditioning the entire decoder output on the latent code.

The decoder utilizes a time-step-wise mask to enforce autoregressive generation. Let $x_1,\ldots$ be the input loops. In predicting the loop $\widehat{y}_i$, at time step $i$, the self-attention layers consider all previous input loops, i.e. loops $x_1$ through $x_{i-1}$, augmented by $x_0=\mathbf{d}$. During training, these previous inputs are teacher-forced, i.e. they are from the ground truth data. During inference time, in the autoregressive scheme, each output $\widehat{y}_i$ is fed back in as the input at the next time step. In both training and inference, the initial embedding $\mathbf d$ is predicted from a latent code $\mathbf z$. 

\subsection{Training and Inference}

We train a VAE to predict the posterior distribution parameters $\color{violet}\mu_q$ and $\color{violet}\log (\sigma_q ^2)$ using the encoder network. 
The latent space is encouraged to behave as a standard Gaussian through (normalized) Kullback-Leibler (KL) divergence, which is given by
\begin{align}
\tilde{L}_\text{KL} &= -\frac12 (1 + {\color{violet}\log (\sigma_q^2)} - {\color{violet}\mu_q^2} - e^{\color{violet}\log (\sigma_q^2)})\nonumber\\
L_\text{KL} &= \frac{\beta_\mathrm{KL}}{N_z} \max(\tilde{L}_\text{KL}, m_\textrm{KL});\label{eqn:klterm}
\end{align}
where hyperparameters $\beta_\text{KL}$ and $m_\text{KL}$ are respectively the KL term weight and minimum. 

The enforcing of this latent space prior has a regularizing effect on reconstruction performance; this trade-off can be tuned via these $\beta_\mathrm{KL}$ and $m_\mathrm{KL}$ hyperparameters. Generally, higher KL weighting implies smoother latent distributions but with poorer variety and decoding fidelity.

% decoder loss stuff:
The reconstruction loss on the decoder output reflects the accuracy of predicting both the loop points and the level-up flag (specifying whether the loop introduces a new plane).
For our loop representation, this implies two terms: the L2 portion (calculated on the raw point coordinates), and the binary cross-entropy portion (calculated on the binary level-up flags). 
Concretely, suppose loops are encoded with $N=32$ points per loop. \rhc{fix this sentence} The embedding for a closed loop is a $2N+1=65$-valued vector $v$, where $v_{[:64]}$ (the first 64 elements of $v$) contains the loop $s_t$, and $v_{[65]}$ (the 65$^\text{th}$ element) is the binary level-up flag $u_t$. Let $\hat y$ be the predicted loop data, and $x$ be the ground truth (input) loop. The reconstruction loss terms are
\begin{align}
    L_R &= L_\text{L2 for coordinates} + L_\text{BCE} \nonumber\\
    &= \left\| \hat y_{[:64]} - x_{[:64]} \right\|^2 +  \nonumber\\&[-(x\cdot \log (\hat y_{[65]}) + (1-x_{[65]}) \cdot \log (1-\hat y_{[65]})] 
    \label{eqn:reco-term}
\end{align}

\subsection{Implementation details}
\paragraph{Dataset preparation}
To prepare a dataset for training with \ourmethod{}, we enforce that all meshes are manifold, normalized to unit cube, and are oriented in a semantically consistent way (i.e. all vases are upright along the $y$-axis; all sofas have their length along the $x$-axis). 
We then select a fixed list of slice planes (we define planes along an axis) whose range and density cover all of the geometry of the input shapes to the extent desired. These slice planes should ideally also be defined such that the same plane should intersect all meshes at roughly the same semantic region, i.e. the first slice plane for a vase dataset should intersect near the bottom of each shape.

\paragraph{End-of-sequence embedding}
Since our decoding is autoregressive, the generated time steps should contain information that determines when the sequence generation should terminate. 

To handle this, we define a special end-of-sequence embedding in the training data to be a loop with all features equal to 0 except for a level-up flag of $1$. In this way, we abort the autoregressive generation when the network-predicted loop points are sufficiently close to zero and the level-up flag is $1$. This is analogous to the \texttt{<EOS>} token in language generation models.

\paragraph{KL weight annealing}
To facilitate better VAE training, we employ an exponential KL weight annealing schedule similar to the scheme used in \cite{ha2018SketchRNN}, in which the weight $\beta_\text{KL}$ of the KL divergence loss term is slowly ramped up to the defined value.

\begin{figure}[h]
    \centering
    \includegraphics[width=0.28\columnwidth]{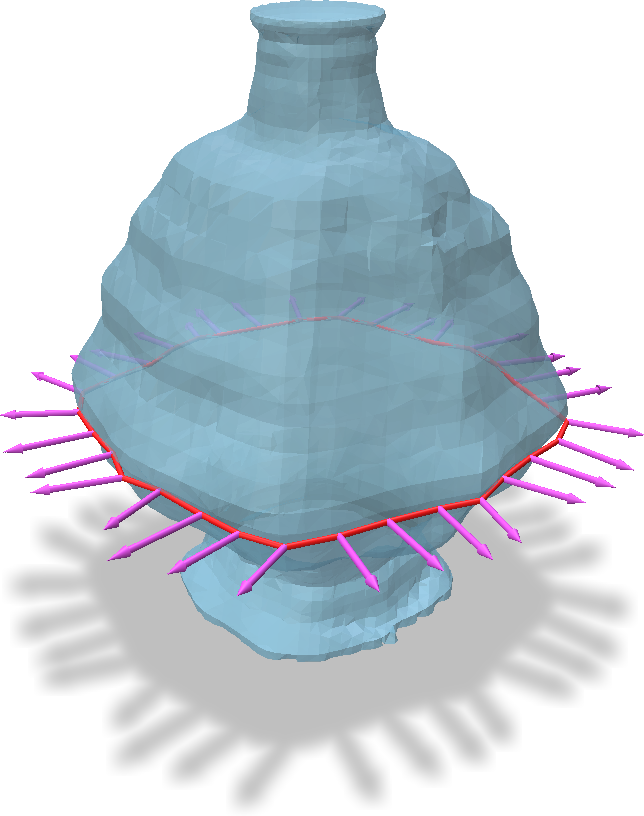}
    \includegraphics[width=0.32\columnwidth]{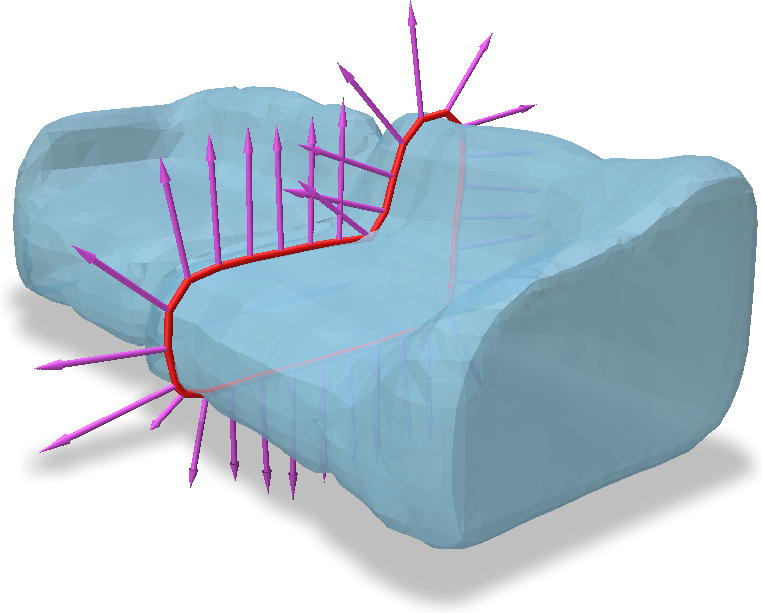}
    \includegraphics[width=0.34\columnwidth]{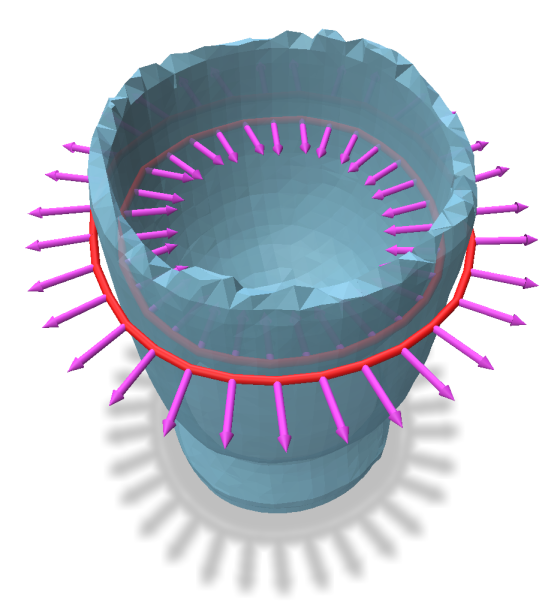}
    \caption{Loop normal estimation. We calculate the loop normal orientation (even for loops contained within other loops, as seen on the right) by leveraging a point known to be outside the loop (e.g., far outside the slice plane).}
    \label{fig:normal-estimation}
\end{figure}
\paragraph{Reconstructing meshes from loops}
All shapes depicted in the figures in our current work are obtained via Poisson reconstruction~\cite{kazhdan2006poisson}, which we found to be the most reliable method for producing shapes that respect the loop hierarchies and the surface they represent.

\begin{figure}[b]
    \centering
    \includegraphics[width=0.82\columnwidth]{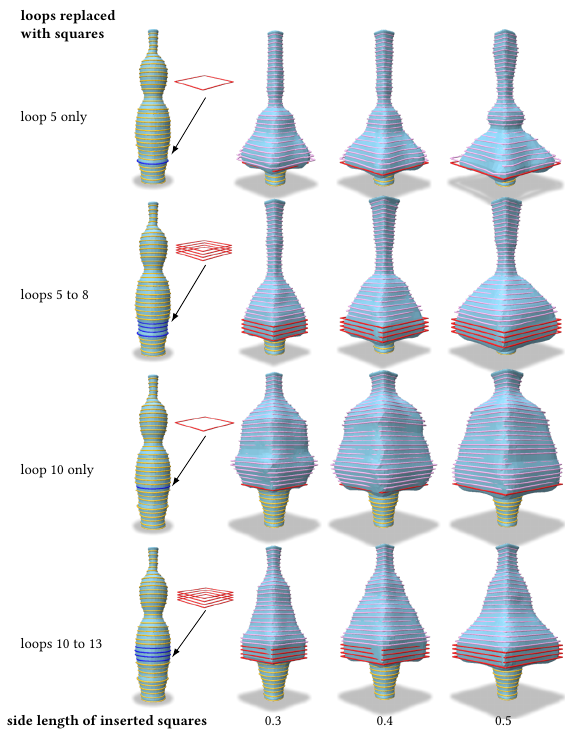}
    ~
    \caption{Injection of loops with square primitives. We manually specify square loops and fix them at certain time steps during autoregressive decoding of a ``recipient shape'' and latent code (untampered original decoding, leftmost shape in each row). When decoding the rest with the edit present, we obtain adaptive shape continuations with sharp angles and natural-looking square cross-sections to match the inserted square loops.}
    \label{fig:square-injection}
\end{figure}

\begin{figure*}
    \newcommand{\putinsetloopfigure}[1]{\frame{\expandafter\includegraphics\expandafter[scale=0.115, trim=0 50 13 100, clip]{#1}}}

    \centering
    \includegraphics[width=\linewidth]{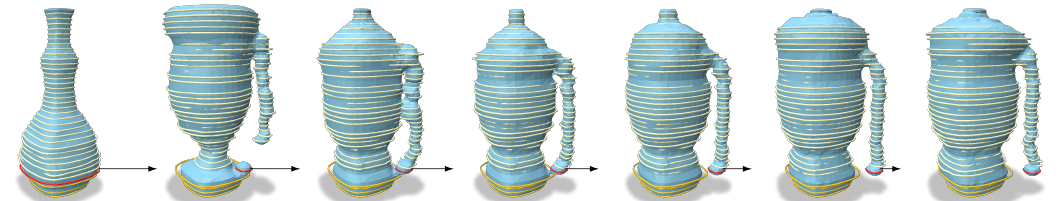}

    % for 01-19 insets
    \putinsetloopfigure{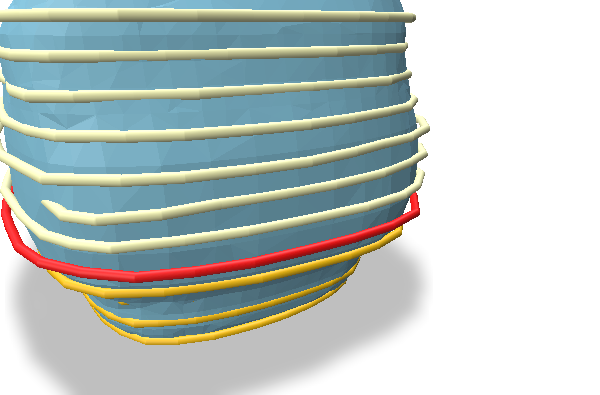}
    \putinsetloopfigure{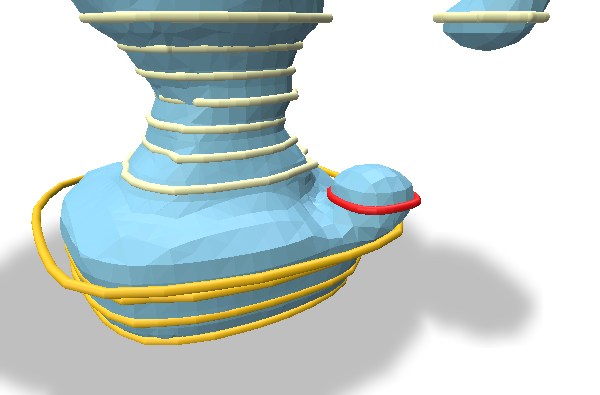}
    \putinsetloopfigure{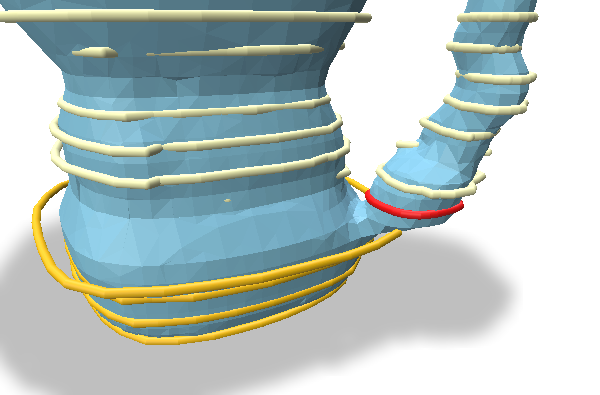}
    \putinsetloopfigure{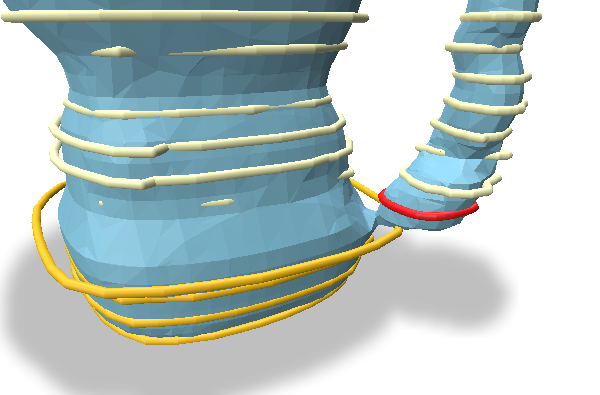}
    \putinsetloopfigure{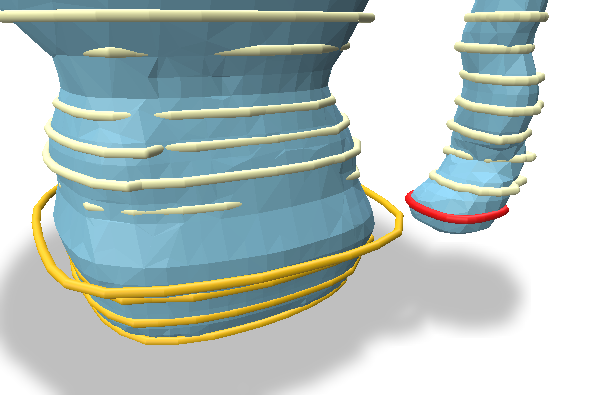}
    \putinsetloopfigure{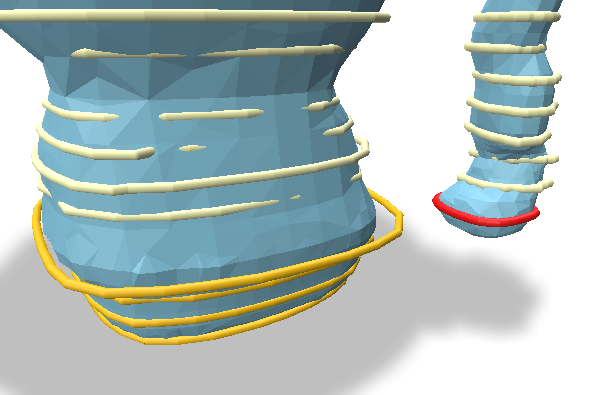}
    \putinsetloopfigure{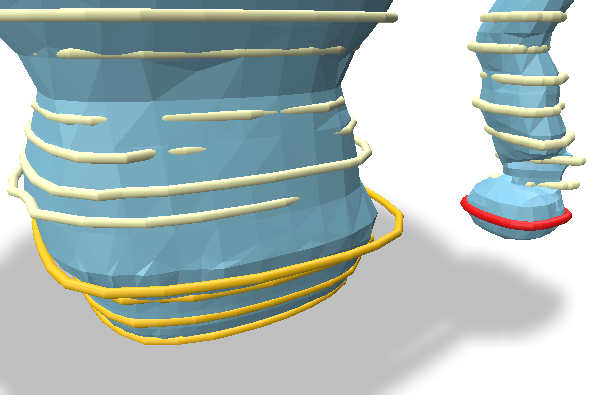}
    
    \caption{Adaptive shape repair and morphing by the manual editing of a single loop during autoregressive decoding. The original untampered shape is the leftmost shape, where the original size and position of the target loop is shown in red. The rest of the shapes show decoding results where the target loop (scaled down by 0.2) is translated away by increasing amounts (see close-ups in inset figures). To best explain the manually-added anomaly of the shifted loop, the model transitions from decoding a symmetrical, handleless vase to decoding a one-handle vase that accommodates the new target loop position.}
    \label{fig:morph-to-mug}
\end{figure*}

\paragraph{Loop-based post-processing heuristics} \label{para:loop-heuristics} The interpretability of the loop representation allows employing loop-based post-processing heuristics to obtain better reconstructions from the raw sequence data. To obtain the necessary normals for Poisson reconstruction, we estimate each loop vertex's normal based on the boundary conditions and line-of-sight. Loops contained within a larger loop in the same plane (representing inner surfaces), can be assigned inwards-facing normals by testing containment of points within the larger contour; this allows us to reconstruct shapes with inner surfaces such as cups and hollow vessels (see \cref{fig:normal-estimation}, right). 
Although such loop-estimated normals will differ from the ground truth surface normals, in practice, Poisson reconstruction robustly reconstructs the underlying surface.

To prevent holes in the reconstruction from an oriented point cloud, we also sample points in the area enclosed by the top and bottom loops; these points' normals are set to be the same as the plane normal (for the topmost loop) or its negative (for the bottom loop). These patches of points form the top and bottom cap (or left and right, if the slice plane axis is horizontal) of each shape when reconstructed.

%% file: 05_experiments_old.tex
\section{Experiments}
In this section, we evaluate our method using several qualitative and quantitative experiments. We demonstrate manual loop editing and intervention possibilities in \cref{sec:editing}. Moreover, we explore the learned latent space generative capabilities and smoothness properties in \cref{sec:latent}. \rh{We propose an alternative neural loop representation for shape editing rather than for purely unconditional generation, thus there is no benchmarks or like-for-like method for direct comparison. As such, we conduct comparisons against existing alternative representations in the supplemental material.}

\label{sec:exp}
We demonstrate our model's capabilities on datasets of two shape categories: vases from COSEG~\cite{wang2012coseg} and sofas from ShapeNet~\cite{shapenet2015}. We chose these due to their regularity and varied topologies (vases) and \nm{angular cross-sectional
features} (sofas).
For each dataset, we fix a set of slice planes that best captures the geometry: the slice planes for COSEG vases are perpendicular to the $y$ axis, and ShapeNet sofas the positive $x$ axis.
The details of the datasets and their slice planes used are in \Cref{tab:dataset-info}. 
In both datasets, we resample every sliced cross-section to $N=32$ vertices per closed polygonal loop. 
We train using the Adam optimizer with a learning rate of $0.00007$ for 70 epochs followed by a linear ramp-down LR schedule for 7230 epochs ending at zero learning rate. \rh{Details about our network architecture can be found in the supplemental material.}

\begin{table}
\caption{Details of the datasets and slice planes used}
\label{tab:dataset-info}
\centering
{\footnotesize 
\begin{tabular}{cccc}
    \toprule
    Dataset & Train size & Test size & \# planes \\ \midrule
    COSEG vases & 231 & 69 & 40 \\
    ShapeNet sofas & 2644 & 529 & 32 \\
     \bottomrule
\end{tabular}
}
\end{table}

\subsection{Loop Editing and Intervention}
\label{sec:editing}

\begin{figure}[b]
    \centering
    \includegraphics[width=0.98\columnwidth]{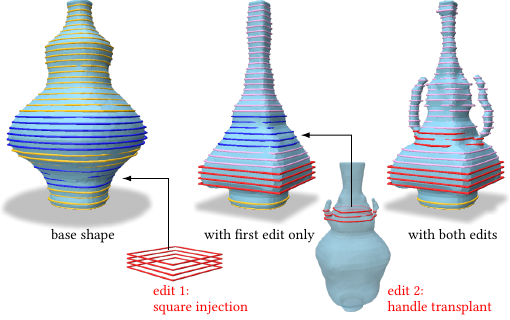}
    ~
    \caption{Compositionality of edits. Two incremental edits: first an injection of square loops, followed by the transplanting of handles from a donor shape. The first edit alone causes the remaining decoded shape to exhibit sharp, right-angled features. Adding onto this intermediate edited shape, we transplant loops belonging to the handles of another vase, causing the growing of handles on the edit target while preserving the square-like cross-sections influenced by the first edit.
    }
    \label{fig:vase-twoedit}
\end{figure}

\begin{figure}
    \centering
    \includegraphics[width=\columnwidth]{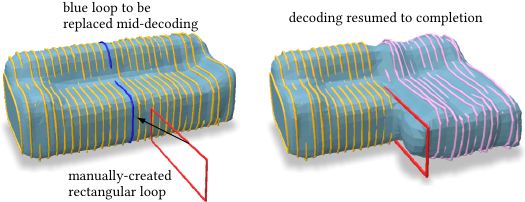}
    ~
    \caption{Injection of loops with rectangular primitives, on sofa dataset. We manually specify a rectangular loop and fix it at a time step during autoregressive decoding of a ``recipient shape'' and latent code (untampered original decoding, first shape). When decoding the rest with the edit present, we obtain an adaptive continuation, an extended sofa seat, to match the inserted long rectangular loop.}
    \label{fig:sofa-rectangle-injection}
\end{figure}

The autoregressive shape prior learned in our decoder through the loop representation grants a multitude of possibilities for interpreting and modifying loop sequences \textit{during} autoregressive decoding to obtain manual control and refinement. 
Here, we present two primary ways of modifying loops---simple transformations, and loop transplantations---during autoregressive decoding. \cref{fig:teaser,fig:handle-transplant,fig:morph-to-mug,fig:square-injection,fig:sofa-rectangle-injection,fig:vase-twoedit} show that combinations of these basic loop modification techniques can reveal properties of our model that allow \textit{controllable} structural modifications, \textit{diverse} geometric reinterpretations, and \textit{intuitive, composable} manipulation.

\subsubsection{In-place modifications to loops}
The first modification method involves translations and scaling on certain loops (i.e. time steps in a sequence) before they are autoregressively appended into the input sequence for the prediction of the next time step. Our model shows an ability to adapt the overall generated shape to best explain such manually-added changes, thereby re-contextualizing user edits and inducing structural modifications in semantically reasonable ways. We first see this in \cref{fig:teaser} where a stack of displaced loops induces a handle.
The new handle is extended accordingly, with the top half of the vase also expanded to accommodate the new protrusion (\cref{fig:teaser}, right). 
% We also see that inserting 3 loops (with smaller handles)
These structural changes are seen in more detail in \cref{fig:morph-to-mug}, where we show continuation shapes resulting from translating one selected loop progressively away from its original location. \nm{Starting from the changes induced by the first edit (the second frame in \cref{fig:morph-to-mug}), subsequent gradual adjustments to the transformation (e.g. translation offset) lead to gradual adaptations which are nonetheless diverse variations on the same motif (e.g. widening variations of the top `lid' as the handle is pulled away from the vessel). Further, such in-place edits still visibly preserve geometric characteristics of the base shape, as shown in \cref{fig:multi-target-handle-pull} which varies the base shape while fixing the handle-pulling edit operation. Such interactions not only induce structural adaptations but also allow a natural design space to be explored via simple loop tweaks by a user.}
% Fig.~\ref{fig:morph-to-mug}
\nmc{don't know if i should say this, because then readers can just ask Where's The User Study}

\subsubsection{Shape blending and composition via loop transplantation}
In this task, we intervene in the autoregressive unfolding of a loop sequence by replacing the generated output at certain time steps with a few manually-provided loops from elsewhere (i.e., taken from another shape's loops as in \cref{fig:handle-transplant}, \cref{fig:vase-twoedit}, or drawn manually as in \cref{fig:square-injection}, \cref{fig:sofa-rectangle-injection}, \cref{fig:vase-twoedit}), and resuming the autoregressive decoding with the edit present.

We demonstrate in \cref{fig:teaser,fig:handle-transplant,fig:square-injection} that this editing method is able to produce meaningful sequence continuations that in effect create ``blend shapes'', re-interpreting the same ``recipient'' shape's latent code but with cross-sectional characteristics from the ``donor'' loops.

\paragraph{Controllable editing via transplant operations}
\cref{fig:square-injection} further demonstrates that various parameters of a transplant such as location (time step) of loop replacement and the transformations applied on the incoming loops enable a diverse, visibly distinct range of shape continuations. As loops are scaled up or down, inserted earlier or later, or injected in more or fewer planes, the adapted shapes are notably not simple rescales or translations of the same result but distinct geometric variants and designs. They offer controllable, natural reinterpretations that best explain both the unedited portion and the new additions.
Finally, \cref{fig:vase-twoedit} shows that edits and insertions of loops are compositional and can be added iteratively, preserving influences of previous edits on the continuation. This composability offers an additional degree of freedom for further controlling the structural modifications induced by loop edits.

These characteristics---shape blending, diverse geometric reinterpretations, and compositionality---offer many avenues for controlling the geometric design of a shape to be completed with \ourmethod{}.

\subsection{Learned Latent Space Properties}
\label{sec:latent}
We demonstrate that a trained model has a smooth latent space that represents a good distribution of shapes, with varying structural features and cross-section styles. 
\cref{fig:interpolation} shows progressions of shapes obtained by decoding latent vectors linearly interpolating between two points in the latent space. We show that interpolation produces intermediate shapes that are sensible with respect to the dataset and smoothly transition between the start and end shapes. 
\cref{fig:random_samples} shows shapes obtained by sampling random latent vectors $\mathbf z$ from
the standard normal distribution (for all models trained on each category, the latent size $N_z$ is $64$.)

%% file: 06_conclusion.tex
\section{Discussion and Future Work}
\begin{figure}[b]
    \centering
    \includegraphics[width=0.8\linewidth]{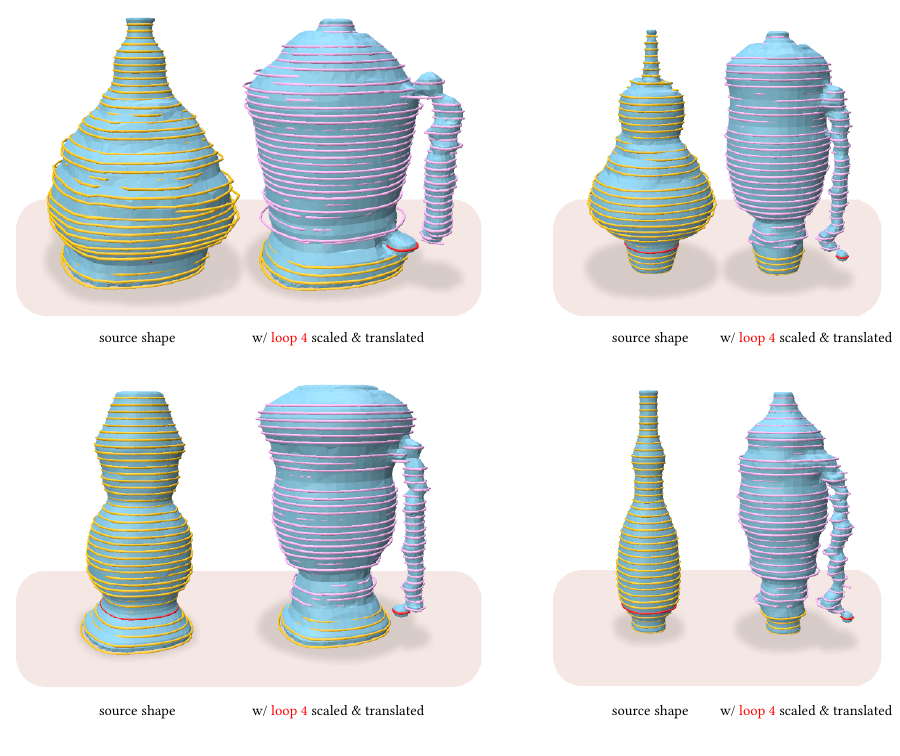}
    \caption{The same edit action---one loop scaled and translated away by the same amount---applied to different base shapes. This same procedure applied to different starting shapes produces one-handle vessels that nonetheless are influenced by the cross-section and body characteristics of the base shape. Thin vases remain thin but with a handle; stout vases likewise give stouter result shapes.}
    \label{fig:multi-target-handle-pull}
\end{figure}

We presented \ourmethod{}, an alternative neural representation for synthesizing and editing shapes based on a sequence of cross-sectional contours. This representation is light-weight, flexible, and capable of representing surfaces with varied geometric structures. Since loops are \emph{explicit}, they can be directly encoded into latent vectors using the proposed \emph{loop-encoder}. Our autoregressive decoder produces a series of sequential loops that are interpretable and intuitively editable, which altogether describe the entire shape.

A notable property of our technique lies in the ability to apply simple yet intuitive transformations to loops. Loop edits imply broader, non-local changes to the resultant geometry via autoregressive decoding that adapts to the modified sequence. We showed that editing techniques applied in various combinations create an editing paradigm featuring controllable structural modifications, diverse design possibilities, and intuitive manipulation.

Currently, our system only considers cross-sectional contours along a single axis; however, an interesting direction for future work is to slice and fuse loops across multiple axes. In addition, not all shape categories possess a natural axis and direction in which to scan slice planes; alternative formulations that treat sets of closed loops as unordered in space may be more appropriate for such situations. Finally, we are also interested in exploring other neural applications of loops, such as for surface reconstruction.

\begin{figure}[]
    \centering
    \includegraphics[width=0.87\columnwidth]{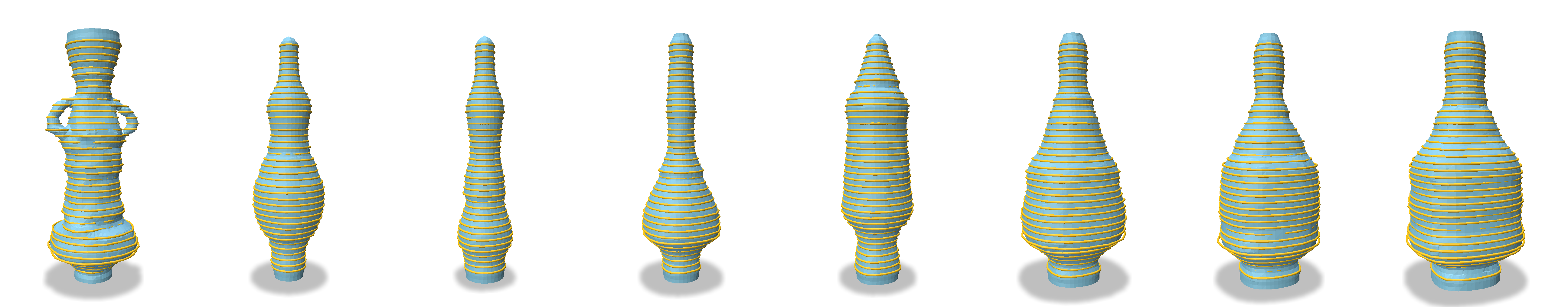}
    \includegraphics[width=0.87\columnwidth]{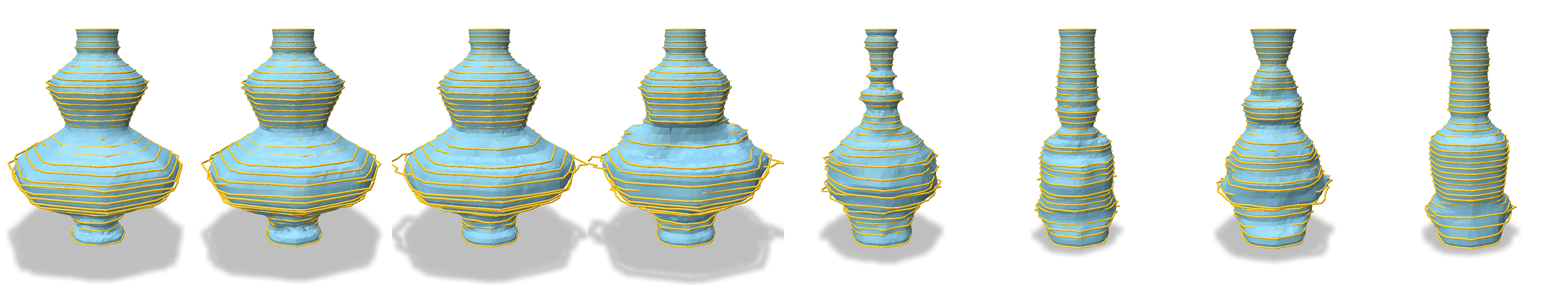}    
    \includegraphics[width=0.86\columnwidth]{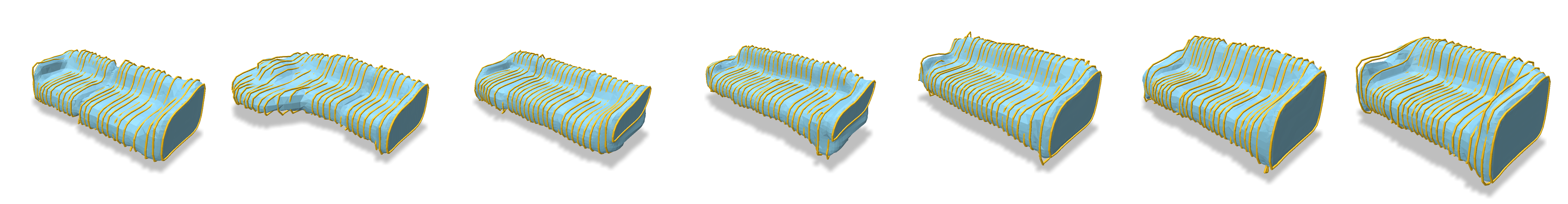}
    \includegraphics[width=0.86\columnwidth]{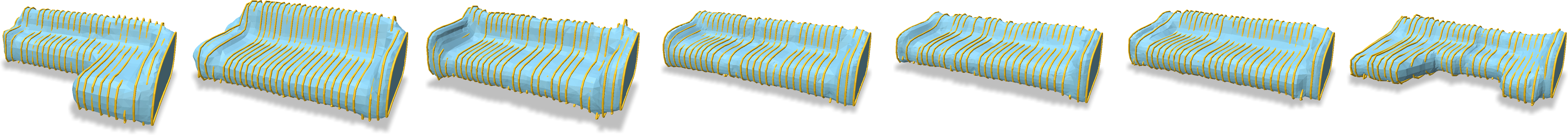}
    \caption{Latent space interpolation for \ourmethod{} is smooth, while still preserving geometric features and changes in structure.}
    \label{fig:interpolation}
\end{figure}

\begin{figure}[]
    \centering
    \includegraphics[width=.8\columnwidth]{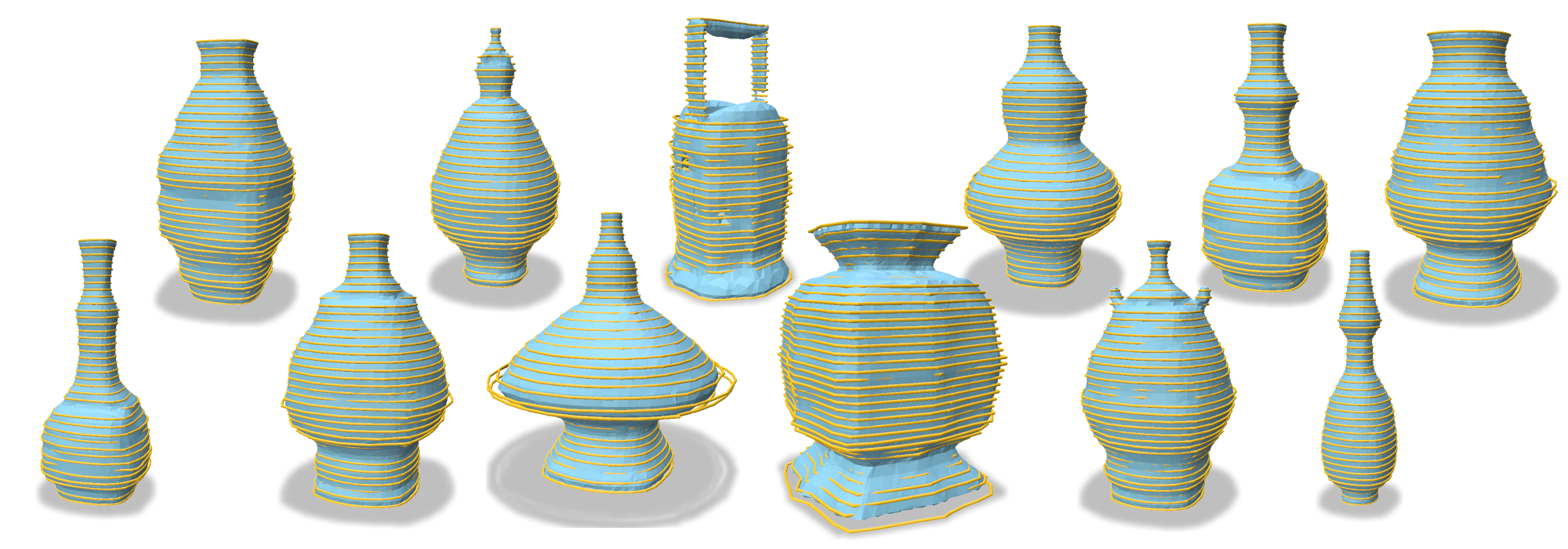}
    \includegraphics[width=.8\columnwidth,trim=0 0 0 80]{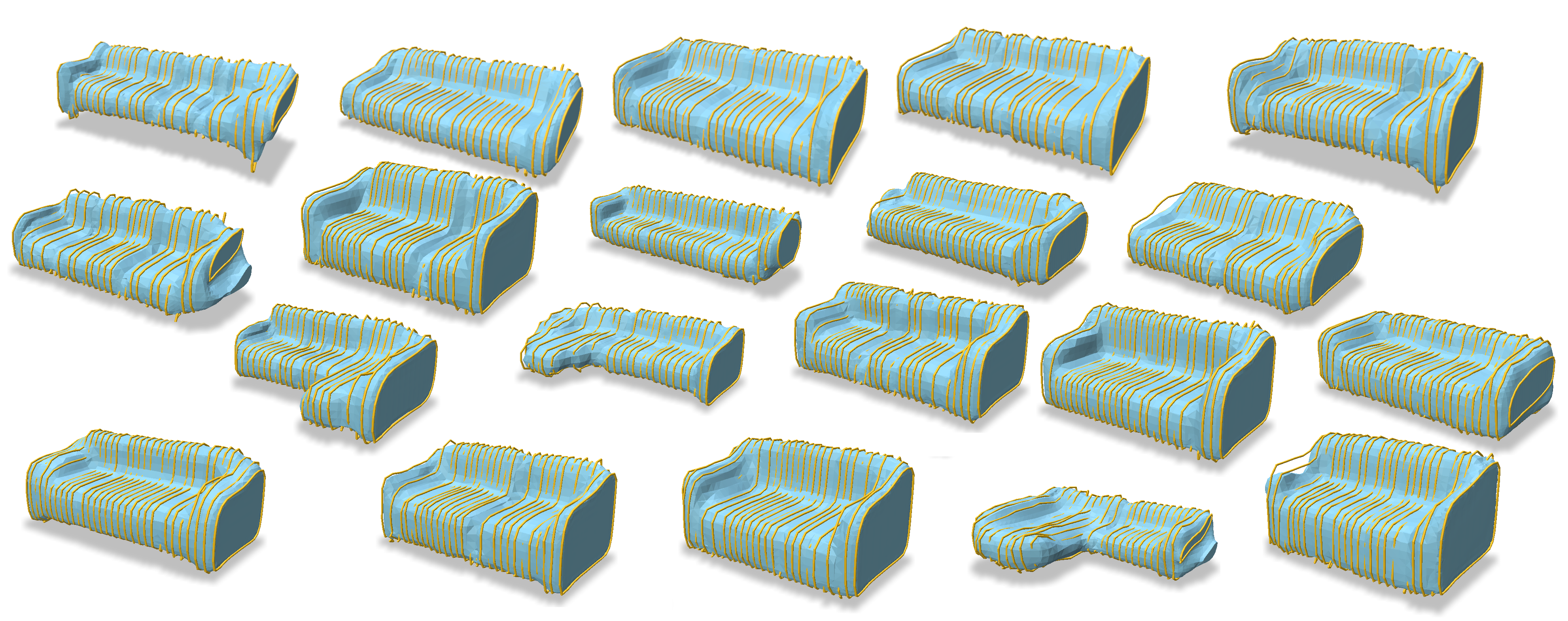}
    \caption{Novel random samples. We sample from the learned latent distribution to generate novel examples which contain varying structural features.}
    \label{fig:random_samples}
\end{figure}

%% file: 08_ack.tex
\section{Acknowledgements}
This work was supported in part by the TRI University 2.0 program, the AFOSR MADlab Center of Excellence, as well as gifts from Adobe Research. We are grateful for the AI cluster resources, services, and staff expertise at the University of Chicago. We thank Xiao Zhang and the members of PALS and 3DL for their helpful comments, suggestions, and insightful discussions.